\documentclass{article}
\usepackage{ijcai25}

\usepackage{amssymb}
\usepackage{multirow}
\usepackage{array}
\usepackage{adjustbox}

\usepackage{times}
\usepackage{soul}
\usepackage{url}
\usepackage[hidelinks]{hyperref}
\usepackage[utf8]{inputenc}
\usepackage[small]{caption}
\usepackage{graphicx}
\usepackage{amsmath}
\usepackage{amsthm}
\usepackage{booktabs}
\usepackage{algorithm}
\usepackage{algorithmic}
\usepackage[switch]{lineno}

\title{Data-Efficient Model for Psychological Resilience Prediction based on Neurological Data}

\author{
Zhi~Zhang$^1$\and
Yan~Liu$^{1,}$\thanks{Corresponding authors.}\and
Mengxia~Gao$^2$\and
Yu~Yang$^3$\and
Jiannong~Cao$^1$\and\\
Wai~Kai~Hou$^3$\and
Shirley~Li$^2$\and
Sonata~Yau$^1$\and
Yun~Kwok~Wing$^4$\And
Tatia~M.~C.~Lee$^{2,\ast}$\\
\affiliations
$^1$The Hong Kong Polytechnic University,
$^2$The University of Hong Kong,
$^3$The Education University of Hong Kong,
$^4$The Chinese University of Hong Kong\\
\emails
zhi271.zhang@connect.polyu.hk,
yan.liu@polyu.edu.hk,
mgao@hku.hk,
yangyy@eduhk.hk,
jiannong.cao@polyu.edu.hk,
wkhou@eduhk.hk,
shirleyx@hku.hk,
sonata.yau@polyu.edu.hk,
ykwing@cuhk.edu.hk,
tmclee@hku.hk
}

\begin{document}

\maketitle

\begin{abstract}
    Psychological resilience, defined as the ability to rebound from adversity, is crucial for mental health. Compared with traditional resilience assessments through self-reported questionnaires, resilience assessments based on neurological data offer more objective results with biological markers, hence significantly enhancing credibility. This paper proposes a novel data-efficient model to address the scarcity of neurological data. We employ Neuro Kolmogorov-Arnold Networks as the structure of the prediction model. In the training stage, a new trait-informed multimodal representation algorithm with a smart chunk technique is proposed to learn the shared latent space with limited data. In the test stage, a new noise-informed inference algorithm is proposed to address the low signal-to-noise ratio of the neurological data. The proposed model not only shows impressive performance on both public datasets and self-constructed datasets but also provides some valuable psychological hypotheses for future research.
\end{abstract}

\section{Introduction}
\label{sec:introduction}
In contemporary society, psychological stress has emerged as a pervasive challenge affecting individuals across demographic boundaries, regardless of age or gender \cite{panicker2019survey}. As research into sources of stress, such as those found in the workplace, has advanced, more researchers realize that eliminating these stressors is challenging \cite{lazarus2020psychological}. Consequently, the study of resilience, describes the maintenance of stable good mental health or the quick recovery of mental health during or after stressor exposure, has garnered increasing attention in recent years \cite{kalisch2017resilience}. Related work shows that after experiencing stressful events, some individuals adapt by maintaining stability and functioning well, whereas others may continue to experience distress \cite{dong2018stress}. 

In the past decades, psychological resilience has been assessed through self-reported questionnaires \cite{connor2003development}, and recent advances in neuropsychology have found the correlations between resilience and neurological signals \cite{lau2021integrative,krause2023predicting,xiang2024using}. For example, the research in functional magnetic resonance imaging (fMRI) research has revealed associations between resilience and specific brain regions \cite{kong2015neural,keynan2019electrical} while electroencephalography (EEG) studies have identified meaningful relationships between resilience and brain networks \cite{paban2019psychological}. Compared to questionnaires, predicting resilience using fMRI and EEG offers two significant advantages: early diagnosis and result reliability. For instance, individuals typically need to first suspect they have low resilience before using specific questionnaires for confirmation, such as the Connor-Davidson Resilience Scale (CD-RISC). In contrast, biological markers for static state can automatically identify individuals with low resilience during routine checks. Additionally, biological markers provide more objective measurements, which are especially valuable in high-risk professions where individuals are continuously exposed to stressors. For example, when assessing pilots' resilience, repeated use of questionnaires may become ineffective due to familiarity with the questions or overly optimistic self-reporting.

This paper addresses a compelling problem: predicting participants' resilience solely based on static state fMRI and EEG data. Despite some insightful findings from previous works, the accuracy of resilience prediction remains unsatisfactory due to the lack of a strong correlation between individual neurological features and resilience values \cite{bonanno2021resilience}. The challenge is further compounded by the scarcity of neuroimaging data, which prevents the direct application of existing multimodal learning techniques that have successfully uncovered complex relationships and complementary information across different modalities in large datasets \cite{gao2022pyramidclip,lee2022uniclip,goel2022cyclip,xue2023ulip}.

To tackle these challenges, we propose a novel data-efficient psychological resilience prediction model incorporating several innovative techniques. Given the limited data, we select Kolmogorov-Arnold Networks (KAN) \cite{liu2024kan,liu2024kan} as the backbone due to their strengths in generalization and interpretability \cite{somvanshi2024survey}. We introduce a new network structure, neuro-KAN, tailored to our multimodal data. Additionally, we enhance the training data volume using a smart chunking algorithm, inspired by the psychological insight that resilience is a relatively stable character trait \cite{steyer1999latent}. The chunking algorithm comes from the simple idea that we partition the collected data into chunks as much as we can while meeting a prerequisite that each chunk is long enough to capture the complete characteristics of stress resilience. For example, if we want to count the heart rate, a 0.1-second chunk is too short to capture a full cardiac cycle while a 24-hour chunk is impractical for diagnosis; clinical practice often uses a 1-minute chunk based on extensive experience. The challenge in our context is determining the appropriate chunk size for fMRI and EEG data for resilience detection. We propose using tree-structured Parzen Estimators to estimate the chunk size while simultaneously learning the shared latent space of the multimodal data. To enhance the alignment accuracy of EEG and fMRI, participant behavior data supervises the learning of the shared latent space in the training stage. To address the low signal-to-noise ratio of neurological data (e.g., muscle artifacts \cite{lai2018artifacts}), we introduce a new noise-informed inference algorithm in test stage, which predicts resilience based on the data quality of EEG and fMRI.

\section{Related Work}
\label{sec:related_work}

\subsection{Kolmogorov-Arnold Networks}

Kolmogorov-Arnold Networks (KAN) \cite{liu2024kan,liu2024kan} are neural networks inspired by the Kolmogorov-Arnold representation theorem, which represents multivariate continuous functions as sums of univariate continuous functions. By introducing learnable univariate functions instead of fixed activation functions, KAN enhances model flexibility and interpretability \cite{somvanshi2024survey}.

Recent studies have demonstrated that KANs offer competitive performance in terms of generalization. Zhang et al. \cite{zhang2024generalization} provide an analysis of KAN's generalization abilities, establishing theoretical generalization bounds. Alter et al. \cite{alter2024robustness} analyze the robustness of KANs under adversarial attacks. Samadi et al. \cite{samadi2024smooth} reveal that smooth KANs embedded with domain-specific knowledge can reduce the data needed for training.
 
Recent applications further show significant potential in health and medical \cite{tang20243d,jahin2024kacq,aghaomidi2024ecg}. However, the application of KANs to EEG and fMRI remains unexplored.

\subsection{Multimodal Representation Learning}

Multimodal representation learning aims to learn unified representations from different data modalities, which has achieved remarkable success across various domains \cite{manzoor2023multimodality}.

Recent advances demonstrate the effectiveness of joint visual-textual embeddings.  Radford et al. propose CLIP \cite{radford2021learning}, and show the alignment between visual and textual features through pre-training on large-scale image-text pairs, enabling zero-shot classification of unseen objects. Subsequent works enhanced this paradigm. Gao et al. \cite{gao2022pyramidclip} propose to align visual and linguistic elements in a hierarchical form. Lee et al. \cite{lee2022uniclip} integrate contrastive losses across multiple domains into a single universal space for improved data efficiency. Goel et al. \cite{goel2022cyclip} optimize representations for geometric consistency across modalities. Beyond image-text pairs, Xu et al. \cite{xue2023ulip} recently extend this approach to three modalities by first learning a common visual-textual space and then aligning it with 3D representations.

While these approaches show promise in domains with abundant data, psychological resilience prediction faces unique challenges due to limited dataset availability, necessitating specialized multimodal learning strategies for small-scale datasets.

\begin{figure*}[]
    \begin{center}
        \includegraphics[width=\linewidth]{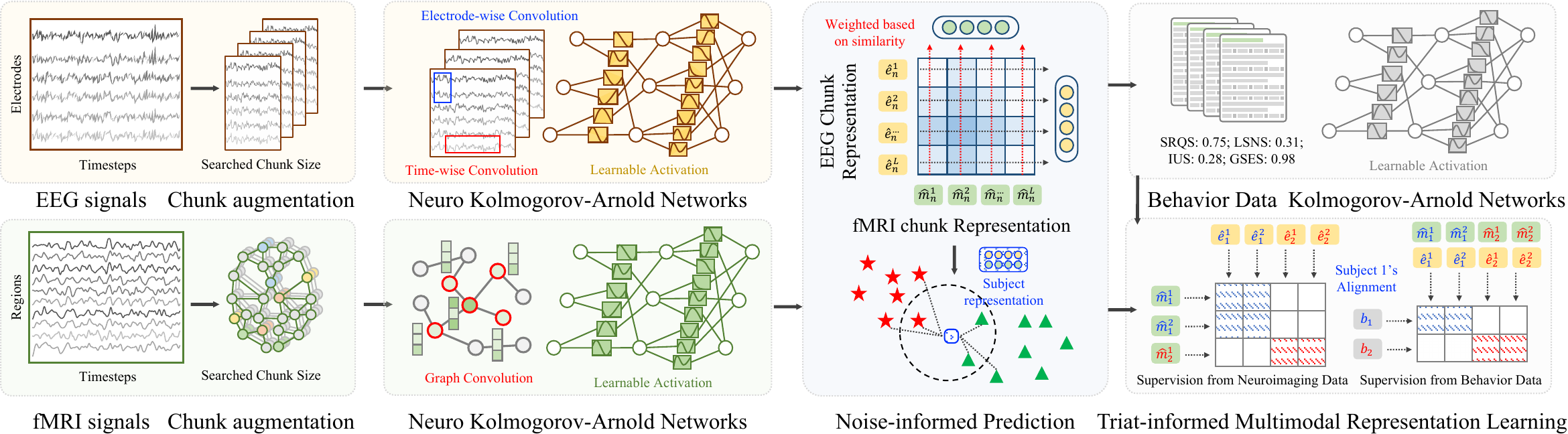}
    \end{center}
    \caption{The overall framework of the data-efficient model for psychological resilience prediction.}
    \label{figure:framework}
\end{figure*}

\section{Methodology}
\label{sec:methodology}

In this section, we propose a method for psychological resilience prediction from multimodal neuroimaging data. Let $\mathcal{D}=\{(e_n, m_n, y_n)\}_{n=1}^N$ denote a dataset of $N$ subjects, where $y_n$ represents the psychological resilience label, and neuroimaging data consists of two modalities. Electroencephalography (EEG) signals $e_n \in \mathbb{R}^{C \times T}$, where $C$ denotes the number of channels and $T$ represents the number of timepoints. Functional magnetic resonance imaging (fMRI) signals $m_n \in \mathbb{R}^{R \times L}$, where $R$ denotes the number of regions of interest (ROIs) and $L$ represents the number of scans.

\subsection{Neuro Kolmogorov-Arnold Networks}
Recognizing the distinct characteristics of each modality, instead of using a single network architecture to process all modalities, we introduce modality-specific variants to KANs.

For EEG signals, which offer excellent temporal resolution capable of capturing neural events at millisecond scale \cite{sturzbecher2012simultaneous}, we propose a hybrid architecture combining KANs with time-wise convolutions and electrode-wise convolutions. Here, we first employ time-wise convolution to model temporal patterns:
\begin{equation}
TWC(X^{TW})_{l,d} = \sigma(\sum_{k=1}^K X^{TW}_{l+k-1,d} \cdot W^{TW}_{k,d})
\end{equation}
where $X^{TW} \in \mathbb{R}^{L \times D}$ represents the input to time-wise convolution, $W^{TW} \in \mathbb{R}^{K \times D}$ denotes learnable temporal filters with kernel size $K$, and $\sigma$ is the SELU activation function \cite{klambauer2017self}. Following time-wise convolutions, we apply electrode-wise convolution to aggregate temporal patterns across electrodes:
\begin{equation}
    EWC(X^{EWC})_{l,o} = \sigma(\sum_{d=1}^D X^{EWC}_{l,d} \cdot W^{EW}_{d,o})
\end{equation}
where $X^{EWC} \in \mathbb{R}^{L \times D}$ represents the input to $EWC$ and $W^{EW} \in \mathbb{R}^{D \times O}$ represents learnable spatial filters that map from $D$ input electrodes to $O$ output channels, with each filter learning to detect specific spatial patterns.

Then, we stack time-wise convolution to recognize patterns through hierarchical receptive fields. To aggregate information, we apply average pooling across timesteps:
\begin{equation}
TP(X^{TP})_{o} = \frac{1}{L} \sum_{l=1}^L X^{TP}_{l,o}
\end{equation}
where $X^{TP} \in \mathbb{R}^{L \times O}$ represents the input to average pooling. Then, we apply KAN to learn nonlinear patterns in EEG signals. The KAN operation is defined as:
\begin{equation}
    KAN(X^{KA})=W^B \sigma(X^{KA})+W^S \operatorname{spline}(X^{KA})
    \label{eq:kan}
\end{equation}
where $X^{KA} \in \mathbb{R}^O$ is a 1D vector serving as input to KAN, $\sigma$ is the SELU activation function, $\operatorname{spline}(X^{KA})=\sum_{i=1}^M c_i B_i(X^{KA})$ represents a B-spline function with $M$ basis functions $B_i$ and learned coefficients $c_i$, and $W^B, W^S \in \mathbb{R}^{D \times O}$ are learnable weight matrices that map to a $D$-dimensional latent space. The final EEG representation can be derived by the output of the KAN operation.

fMRI provides complementary characteristics with superior spatial resolution at the millimeter scale \cite{sturzbecher2012simultaneous}. To leverage this spatial precision, we propose to combine KANs with graph convolution networks \cite{kipf2017semi} to model the complex interactions between brain regions. We represent each fMRI as a functional connectivity network. We use the Shen 268 \cite{shen2013groupwise} node atlas to define regions of interest (ROIs) as nodes. Then, we calculate Pearson correlation coefficients between mean time series in ROIs as functional connections, assigned as edges. Finally, the network can be represented as an adjacency matrix $A \in \mathbb{R}^{R \times R}$, where $R$ denotes the number of ROIs. Each node $i$'s feature is a feature vector $x_i \in \mathbb{R}^R$, from the $i$-th row of $A$. The graph convolution operation then can be defined as:
\begin{equation}
    GCON(X^{GC}) = \alpha(\tilde{D}^{-\frac{1}{2}}\tilde{A}\tilde{D}^{-\frac{1}{2}}X^{GC}W^{GC})
\end{equation}
where $\tilde{A} = A + I_R$ is the adjacency matrix with added self-connections ($I_R$ is the identity matrix), derived from the functional connectivity matrix $A$, $\tilde{D}_{ii}=\sum_j \tilde{A}_{ij}$ is the corresponding degree matrix, $X^{GC} \in \mathbb{R}^{R \times R}$ represents the node feature matrix, $W^{GC} \in \mathbb{R}^{R \times F}$ is the learnable weight matrix mapping to $F$ output features, and $\alpha$ is the LeakyReLU activation function.

To aggregate spatial information, we apply average pooling across all ROIs:
\begin{equation}
RP(X^{RP}) = \frac{1}{R} \sum_{i=1}^R X^{RP}_{i}
\end{equation}
where $R$ denotes the total number of ROIs, and $X^{RP}_{i} \in \mathbb{R}^F$ represents the learned feature vector of the $i$-th ROI after graph convolution. Then, we apply KAN to learn nonlinear relationships in the pooled fMRI representations following Eq. \ref{eq:kan}.

\subsection{Triat-informed Multimodal Representation Learning}

Recognizing that resilience is a relatively stable character trait \cite{steyer1999latent}. For example, one hour or two hours after falling asleep, it is rare for subjects to show significant increase or decrease in stress resilience. Meanwhile, neuroimaging data is typically recorded over long periods, such as sleep EEG which is usually recorded for around 6 hours. We propose to chunk the long recording of neuroimaging data into small time segments, and use these chunks as samples. In detail, we define a chunk augmentation function $AUG(\cdot)$ as:
\begin{equation}
    \begin{aligned}
    AUG({X})=\{{x}_i \mid &{x}_i = \{{X}_{d,t} \mid d \in[1, D], \\
    &t \in[i S, i S+L]\}\}_{i=0}^{K-1}
    \end{aligned}
    \label{eq:augmentation}
\end{equation}
where $L$ denotes the chunk length, $V$ denotes the overlap between adjacent chunks, and $S=L-V$.

Because both modalities reflect resilience-related neural patterns and resilience is a stable trait, we hypothesize that resilience-related representations from the same subject should maintain consistency across both temporal chunks and modalities. Therefore, we propose contrastive learning to align multimodal representations. Following CLIP \cite{radford2021learning}, we use the batch of data to construct positive and negative pairs, training the model to maximize the similarity between positive pairs and minimize the similarity between negative pairs:

\begin{equation}
    \begin{aligned}
    \mathcal{L}_{neu} = \frac{1}{2}(\mathcal{L}(E, M) + \mathcal{L}(M, E))
    \label{eq:contrastive_neu}
    \end{aligned}
\end{equation}
where $E$ and $M$ represent batches of EEG and fMRI features respectively.

While CLIP treats an image and its paired text as positive pairs and other combinations as negative pairs, directly applying this strategy would lead to false negative pairs, as EEG and fMRI chunks sampled from different time segments of the same subject share the same resilience trait. Thus, we propose to treat EEG-fMRI pairs from the same subject as positive pairs, and sample negative pairs from other subjects:
\begin{equation}
    \begin{aligned}
    \mathcal{L}(A,B) = -\frac{1}{|A|} \sum_{a_i \in A} \log \frac{\sum_{b_j \in B_{I(i)}} \exp(\operatorname{sim}(a_i, b_j)/\tau)}{\sum_{b_j \in B} \exp(\operatorname{sim}(a_i, b_j)/\tau)}
    \end{aligned}
    \label{eq:contrastive}
\end{equation}
where $B_{I(i)}$ denotes the set of samples in $B$ from the same subject as $a_i$, $\operatorname{sim}(\cdot,\cdot)$ is the cosine similarity, and $\tau$ is a temperature parameter.

Learning subject-dependent information by matching EEG and fMRI from the same subject is insufficient when the test set contains unseen subjects. It is requried to learn psychological state patterns independent of individual subjects to enable generalization to unseen subjects. Thus, we incorporate behavioral data as supervision during training. We collect psychological scales focusing on social relationships, meaning of life, and other factors demonstrated by prior research to be relevant to resilience mechanisms. We denote the behavioral data as $h_n \in \mathbb{R}^{Q}$, where $Q$ represents the total number of scales. We first use Eq. \ref{eq:kan} to map $h_n$ to non-linear space. Then, we supervise the EEG and fMRI models to extract behavior-related information. We frame this as another multimodal alignment problem, aligning behavior representations with EEG and fMRI representations.

\begin{equation}
    \begin{aligned}
    \mathcal{L}_{beh} = \frac{1}{2}(\mathcal{L}(E, H) + \mathcal{L}(M, H))
    \end{aligned}
    \label{eq:contrastive_beh}
\end{equation}
where subjects with similar psychological states have similar behavioral data, and the neuroimaging data matched with similar behavior data should also be similar.

In different neuroimaging data recording, we find the optimal chunk size is not the same. Chunks that are too short may not contain sufficient information for resilience detection, while excessively long chunks limit the number of samples. To tackle this problem, we further conduct leave-one-subject-out cross-validation on the training dataset. We record the average performance of validation subjects, then use tree-structured Parzen estimators \cite{watanabe2023tree} to determine the optimal chunk length. Mathematically, the optimal chunk length can be defined as:

\begin{equation}
\begin{aligned}
L_{\text{opt}} = \arg \max_{L} \sum_{n=1}^{N} \mathcal{P}(y_n, \hat{y}_n)
\end{aligned}
\label{eq:optimal}
\end{equation}
where $\mathcal{P}$ denotes the performance metric, $y_n$ represents the ground-truth resilience label of the $n$-th subject, and $\hat{y}_n$ represents the predicted resilience label of the $n$-th subject on the validation dataset.

\subsection{Noise-informed Psychological Resilience Prediction}

It is known that EEG and fMRI can contain noise unrelated to resilience. For example, eye movements in EEG signals can affect signal quality \cite{jiang2019removal}. Physiological changes can lead to fMRI signal variations that represent noise rather than brain activity of interest \cite{liu2016noise}.

We observe that while both EEG and fMRI data may be affected by noise, causing out-of-distribution representations for psychological resilience, they are influenced by different noise sources, resulting in distinct distributions. When both EEG and fMRI have low noise levels, their in-distribution psychological resilience patterns lead to similar representations, as they successfully capture subject patterns enforced by the contrastive loss without noise interference.

Therefore, we propose to calculate the similarity between EEG and fMRI representations of the same subject and use this similarity as a metric to identify noisy chunks, assigning them lower weights:

\begin{equation}
    \begin{aligned}
    SIM(n)=&\Big[\sum_{i} \operatorname{sim}\left(\hat{e}_n^i, \hat{m}_n^j\right) \hat{m}_n^j \| \sum_{j} \operatorname{sim}\left(\hat{e}_n^i, \hat{m}_n^j\right) \hat{e}_n^i\Big]
    \end{aligned}
    \label{eq:mil}
\end{equation}
where $\hat{e}_n^{i}$ and $\hat{m}_n^{j}$ denote the representations from the $i$-th chunk of EEG and $j$-th chunk of fMRI of subject $n$ respectively, and $\|$ represents concatenation. Finally, we can use a simple k-nearest neighbors approach to derive new subjects's prediction by retrieving the most similar subjects:
\begin{equation}
y = \frac{1}{k}\sum_{i \in \mathcal{N}_k(SIM(n))} y_i
\end{equation}
where $\mathcal{N}_k(SIM(n))$ denotes the indices of k-nearest neighbors of $SIM(n)$ in the training set representations.
\begin{table*}
    \caption{Comparative experiments of our proposed method versus existing EEG models, fMRI models, and multimodal learning approaches on RESIL and LEMON datasets}
    \begin{center}
      \begin{adjustbox}{width=\textwidth}
          \begin{tabular}{|cl|rrrr|rrrr|}
              \hline
              \multicolumn{2}{|c|}{\multirow{2}{*}{Method}}                                        & \multicolumn{4}{c|}{RESIL Dataset}                                                                                                           & \multicolumn{4}{c|}{LEMON Dataset}                                                                                                           \\ \cline{3-10} 
              \multicolumn{2}{|c|}{}                                                               & \multicolumn{1}{c|}{Acc.$\uparrow$}            & \multicolumn{1}{c|}{F1$\uparrow$}              & \multicolumn{1}{c|}{MAE$\downarrow$}             & \multicolumn{1}{c|}{R2$\uparrow$} & \multicolumn{1}{c|}{Acc.$\uparrow$}            & \multicolumn{1}{c|}{F1$\uparrow$}              & \multicolumn{1}{c|}{MAE$\downarrow$}             & \multicolumn{1}{c|}{R2$\uparrow$} \\ \hline
              \multicolumn{1}{|c|}{\multirow{10}{*}{Model Architectures}}& TCNet \cite{ingolfsson2020eeg}                  & \multicolumn{1}{r|}{0.6667}          & \multicolumn{1}{r|}{0.7170}          & \multicolumn{1}{r|}{0.1537}          & 0.0790                  & \multicolumn{1}{r|}{0.6513}          & \multicolumn{1}{r|}{0.6304}          & \multicolumn{1}{r|}{0.1649}          & 0.1038   \\ \cline{2-10}          
              \multicolumn{1}{|c|}{}                                     & LMDA \cite{miao2023lmda}                        & \multicolumn{1}{r|}{0.7111}          & \multicolumn{1}{r|}{0.7347}          & \multicolumn{1}{r|}{0.1519}          & 0.1714                  & \multicolumn{1}{r|}{0.6513}          & \multicolumn{1}{r|}{0.6344}          & \multicolumn{1}{r|}{0.1638}          & 0.1144   \\ \cline{2-10} 
              \multicolumn{1}{|c|}{}                                     & BNT \cite{kan2022brain}                         & \multicolumn{1}{r|}{0.7333}          & \multicolumn{1}{r|}{0.7667}          & \multicolumn{1}{r|}{0.1487}          & 0.1892                  & \multicolumn{1}{r|}{0.6923}          & \multicolumn{1}{r|}{0.6902}          & \multicolumn{1}{r|}{0.1494}          & 0.2708   \\ \cline{2-10} 
              \multicolumn{1}{|c|}{}                                     & Labram \cite{jiang2024large}                    & \multicolumn{1}{r|}{0.7556}          & \multicolumn{1}{r|}{0.7755}          & \multicolumn{1}{r|}{0.1421}          & 0.2100                  & \multicolumn{1}{r|}{0.6564}          & \multicolumn{1}{r|}{0.6298}          & \multicolumn{1}{r|}{0.1619}          & 0.1477   \\ \cline{2-10} 
              \multicolumn{1}{|c|}{}                                     & CSPNet \cite{jiang2024csp}                      & \multicolumn{1}{r|}{0.7556}          & \multicolumn{1}{r|}{0.7843}          & \multicolumn{1}{r|}{0.1413}          & 0.2285                  & \multicolumn{1}{r|}{0.6769}          & \multicolumn{1}{r|}{0.6358}          & \multicolumn{1}{r|}{0.1578}          & 0.1891   \\ \cline{2-10} 
              \multicolumn{1}{|c|}{}                                     & TSLANet \cite{eldele2024tslanet}                & \multicolumn{1}{r|}{0.7556}          & \multicolumn{1}{r|}{0.7925}          & \multicolumn{1}{r|}{0.1399}          & 0.2315                  & \multicolumn{1}{r|}{0.6769}          & \multicolumn{1}{r|}{0.6519}          & \multicolumn{1}{r|}{0.1562}          & 0.2202   \\ \cline{2-10} 
              \multicolumn{1}{|c|}{}                                     & Conformer \cite{song2022eeg}                    & \multicolumn{1}{r|}{0.7778}          & \multicolumn{1}{r|}{0.8000}          & \multicolumn{1}{r|}{0.1382}          & 0.2500                  & \multicolumn{1}{r|}{0.6974}          & \multicolumn{1}{r|}{0.6911}          & \multicolumn{1}{r|}{0.1470}          & 0.2810   \\ \cline{2-10}  
              \multicolumn{1}{|c|}{}                                     & BrainNPT \cite{hu2024brainnpt}                  & \multicolumn{1}{r|}{0.8000}          & \multicolumn{1}{r|}{0.8235}          & \multicolumn{1}{r|}{0.1371}          & 0.2908                  & \multicolumn{1}{r|}{0.7016}          & \multicolumn{1}{r|}{0.6814}          & \multicolumn{1}{r|}{0.1505}          & 0.2935   \\ \cline{2-10} 
              \multicolumn{1}{|c|}{}                                     & EEGNet \cite{lawhern2018eegnet}                 & \multicolumn{1}{r|}{0.8222}          & \multicolumn{1}{r|}{0.8400}          & \multicolumn{1}{r|}{0.1367}          & 0.3094                  & \multicolumn{1}{r|}{0.7026}          & \multicolumn{1}{r|}{0.6915}          & \multicolumn{1}{r|}{0.1480}          & 0.3084   \\ \cline{2-10} 
              \multicolumn{1}{|c|}{}                                     & BrainGNN \cite{li2021braingnn}                  & \multicolumn{1}{r|}{0.8222}          & \multicolumn{1}{r|}{0.8519}          & \multicolumn{1}{r|}{0.1368}          & 0.2927                  & \multicolumn{1}{r|}{0.7107}          & \multicolumn{1}{r|}{0.6924}          & \multicolumn{1}{r|}{0.1421}          & 0.3398   \\ \hline
              \multicolumn{1}{|c|}{\multirow{5}{*}{Learning Algorithms}} & DCCA \cite{andrew2013deep}                      & \multicolumn{1}{r|}{0.7458}          & \multicolumn{1}{r|}{0.7692}          & \multicolumn{1}{r|}{0.1433}          & 0.1945                  & \multicolumn{1}{r|}{0.6300}          & \multicolumn{1}{r|}{0.6100}          & \multicolumn{1}{r|}{0.1703}          & 0.0817   \\ \cline{2-10} 
              \multicolumn{1}{|c|}{}                                     & DGCCA \cite{benton2019deep}                     & \multicolumn{1}{r|}{0.7750}          & \multicolumn{1}{r|}{0.8085}          & \multicolumn{1}{r|}{0.1379}          & 0.2789                  & \multicolumn{1}{r|}{0.6821}          & \multicolumn{1}{r|}{0.6220}          & \multicolumn{1}{r|}{0.1550}          & 0.2652   \\ \cline{2-10} 
              \multicolumn{1}{|c|}{}                                     & SLIP \cite{mu2022slip}                          & \multicolumn{1}{r|}{0.7778}          & \multicolumn{1}{r|}{0.8148}          & \multicolumn{1}{r|}{0.1357}          & 0.2961                  & \multicolumn{1}{r|}{0.6205}          & \multicolumn{1}{r|}{0.5843}          & \multicolumn{1}{r|}{0.1714}          & 0.0765   \\ \cline{2-10} 
              \multicolumn{1}{|c|}{}                                     & CLIPMixup \cite{oh2024geodesic}                 & \multicolumn{1}{r|}{0.8000}          & \multicolumn{1}{r|}{0.8235}          & \multicolumn{1}{r|}{0.1349}          & 0.3435                  & \multicolumn{1}{r|}{0.6462}          & \multicolumn{1}{r|}{0.6349}          & \multicolumn{1}{r|}{0.1689}          & 0.0952   \\ \cline{2-10} 
              \multicolumn{1}{|c|}{}                                     & CLIP \cite{radford2021learning}                 & \multicolumn{1}{r|}{0.8222}          & \multicolumn{1}{r|}{0.8261}          & \multicolumn{1}{r|}{0.1280}          & 0.3533                  & \multicolumn{1}{r|}{0.7077}          & \multicolumn{1}{r|}{0.6984}          & \multicolumn{1}{r|}{0.1471}          & 0.3234   \\ \hline
              \multicolumn{2}{|c|}{Proposed}                             & \multicolumn{1}{r|}{\textbf{0.8444}}            & \multicolumn{1}{r|}{\textbf{0.8627}} & \multicolumn{1}{r|}{\textbf{0.1203}} & \textbf{0.4185}         & \multicolumn{1}{r|}{\textbf{0.7897}} & \multicolumn{1}{r|}{\textbf{0.7657}} & \multicolumn{1}{r|}{\textbf{0.1455}} & \textbf{0.3510}         \\ \hline
          \end{tabular}
      \end{adjustbox}
    \end{center}
  \label{table:comparasion_experiments}
\end{table*}

\begin{table}
    \caption{Ablation study removing Neuro Kolmogorov-Arnold Networks (KAN), Triat-informed Multimodal Representation Learning (MRL), and Noise-informed Prediction (NP) on RESIL and LEMON datasets}
    \begin{center}
      \begin{adjustbox}{width=0.95\linewidth}
        \begin{tabular}{|ccc|cc|cc|}
            \hline
            \multicolumn{3}{|c|}{Method}                                & \multicolumn{2}{c|}{RESIL Dataset}                     & \multicolumn{2}{c|}{LEMON Dataset}                   \\ \hline
            \multicolumn{1}{|c|}{KAN} & \multicolumn{1}{c|}{MRL} & NP & \multicolumn{1}{c|}{Acc. $\uparrow$}            & F1 $\uparrow$             & \multicolumn{1}{c|}{Acc. $\uparrow$}            & F1 $\uparrow$             \\ \hline
            \multicolumn{1}{|c|}{\scalebox{0.85}[1]{$\times$}}    & \multicolumn{1}{c|}{$\checkmark$}    & $\checkmark$   & \multicolumn{1}{c|}{0.7778}          & 0.8148          & \multicolumn{1}{c|}{0.7692}          & 0.7486          \\ \hline
            \multicolumn{1}{|c|}{$\checkmark$}   & \multicolumn{1}{c|}{\scalebox{0.85}[1]{$\times$}}     & $\checkmark$   & \multicolumn{1}{c|}{0.8222}          & 0.8261          & \multicolumn{1}{c|}{0.7077}          & 0.6984          \\ \hline
            \multicolumn{1}{|c|}{$\checkmark$}   & \multicolumn{1}{c|}{$\checkmark$}    & \scalebox{0.85}[1]{$\times$} & \multicolumn{1}{c|}{0.7773}          & 0.8064          & \multicolumn{1}{c|}{0.7098}          & 0.6888          \\ \hline
            \multicolumn{1}{|c|}{$\checkmark$}   & \multicolumn{1}{c|}{$\checkmark$}    & $\checkmark$   & \multicolumn{1}{c|}{\textbf{0.8444}} & \textbf{0.8627} & \multicolumn{1}{c|}{\textbf{0.7897}} & \textbf{0.7657} \\ \hline
        \end{tabular}
      \end{adjustbox}
    \end{center}
  \label{table:ablation_study}
\end{table}

\section{Experiments}
We conducted experiments on two datasets: the RESIL and LEMON dataset \cite{babayan2019mind}. The RESIL Dataset is a self-constructed dataset comprising 45 subjects (21 males, 24 females, average age 29.9). For EEG data, signals are recorded for each subject during sleep using six electrodes (F3, F4, C3, C4, O1, O2) placed following the 10-20 system. The sleep recordings were bandpass filtered (0.3Hz–35Hz), resampled to 100Hz, and normalized according to the Z-score standardization \cite{wang2024generalizable}. Then, we obtained the imaging data using a 3T Philips MRI scanner, with the resting-state fMRI data acquired and processed following Gao et al. \cite{gao2020connectome}. For behavioral data, we collected the social relationship quality scale (SRQS), Lubben social network scale (LSNS), intolerance of uncertainty scale (IUS), general self-efficacy scale (GSES), and meaning in-life questionnaire (MLQ). We use the Connor-Davidson resilience scale (CD-RISC) as ground-truth.

The LEMON Dataset is a public dataset comprising 227 healthy participants comprising a young (108 males, 45 females, average age 25.1) and an elderly group (37 males, 37 females, average age 67.6).

In the experiments, we trained the model for 20 epochs with a batch size of 128 samples. The learning rate was set to 0.001 with a weight decay of 0.0. For prediction, we set $k$ to 5. We used the Adam optimizer for optimization. We conduct leave-one-subject-out cross-validation to split training and test datasets. We evaluate the proposed method on both classification and regression tasks. We report accuracy, F1 score, Mean Absolute Error (MAE), and R2 score on the test set. 

\subsection{Comparative Experiments}
We first conduct comparative experiments on the RESIL dataset and the LEMON dataset to evaluate the performance of the proposed method against state-of-the-art methods. We first compare our method with multimodal learning methods, including DCCA \cite{andrew2013deep}, DGCCA \cite{benton2019deep}, CLIP \cite{radford2021learning}, SLIP \cite{mu2022slip}, and CLIPMixup \cite{oh2024geodesic}, by replacing the proposed multimodal learning algorithm. We then replace temporal-aware KAN with EEG models, including EEGNet \cite{lawhern2018eegnet}, TCNet \cite{ingolfsson2020eeg}, Conformer \cite{song2022eeg}, LMDA \cite{miao2023lmda}, Labram \cite{jiang2024large}, CSPNet \cite{jiang2024csp}, and TSLANet \cite{eldele2024tslanet}. Finally, we replace spatial-aware KAN with fMRI models, including BrainGNN \cite{li2021braingnn}, BNT \cite{kan2022brain}, and BrainNPT \cite{hu2024brainnpt}.

As shown in Table \ref{table:comparasion_experiments}, the proposed method outperforms existing methods on both datasets and achieves an accuracy of over 0.75 and R2 over 0.3, demonstrating its effectiveness in modeling resilience with multimodal biomarkers. We can find that CLIP-based multimodal learning shows a strong benchmark against other multimodal learning algorithms even on small datasets, but its performance drops on the LEMON dataset where the number of subjects increases and inter-subject variability becomes larger, indicating the necessity of subject-aware design. We also observe that recent large models based on transformers, despite showing impressive performance on large datasets, such as Labram and BrainNPT, are outperformed on small datasets by simpler models based on convolutional networks (e.g., EEGNet) and graph convolutional networks (e.g., BrainGNN). This phenomenon highlights the importance of data-efficient design for small multimodal datasets.

\subsection{Ablation Study}

\begin{figure*}[]
    \begin{center}
    \includegraphics[width=0.95\linewidth]{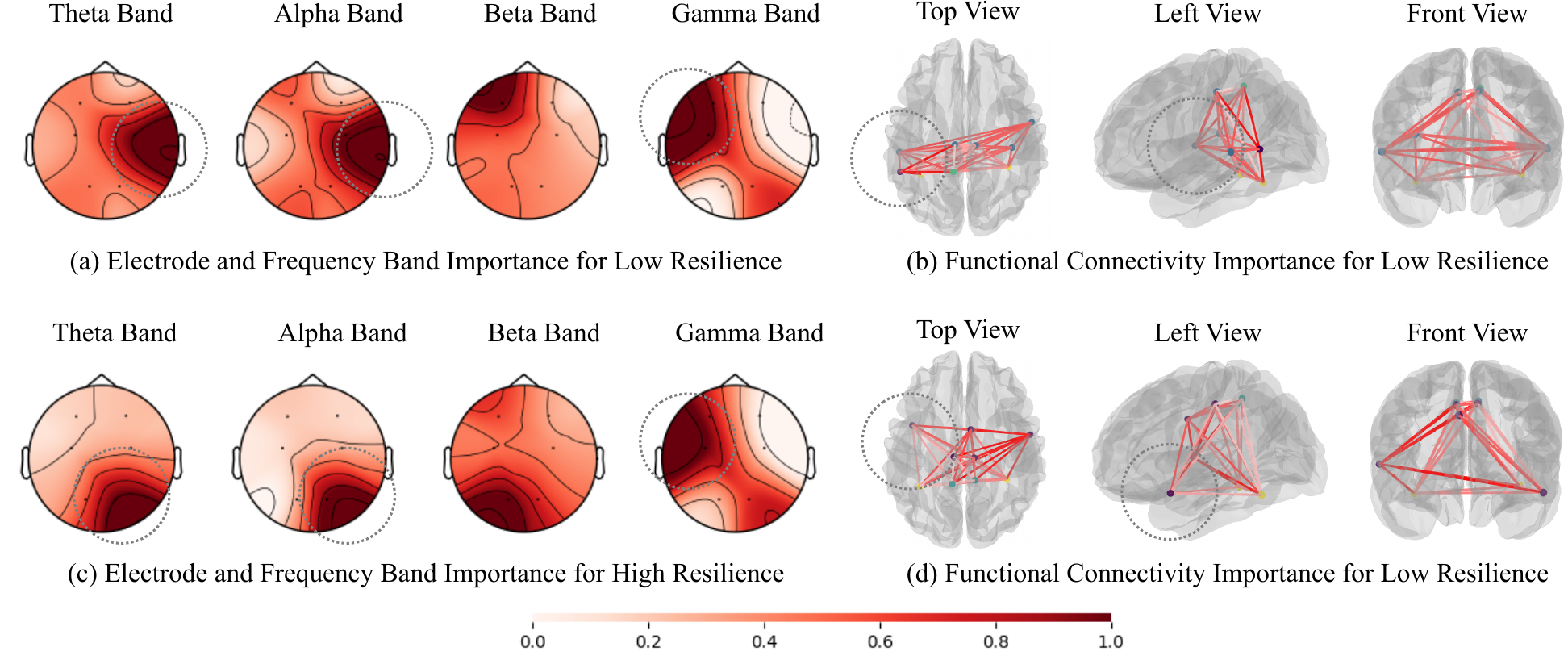}
    \end{center}
    \caption{Interpretability analysis of EEG topography and functional connectivity on the RESIL dataset.}
    \label{figure:eeg_fmri}
    \end{figure*}
    
\begin{figure}[]
    \begin{center}
    \includegraphics[width=0.75\linewidth]{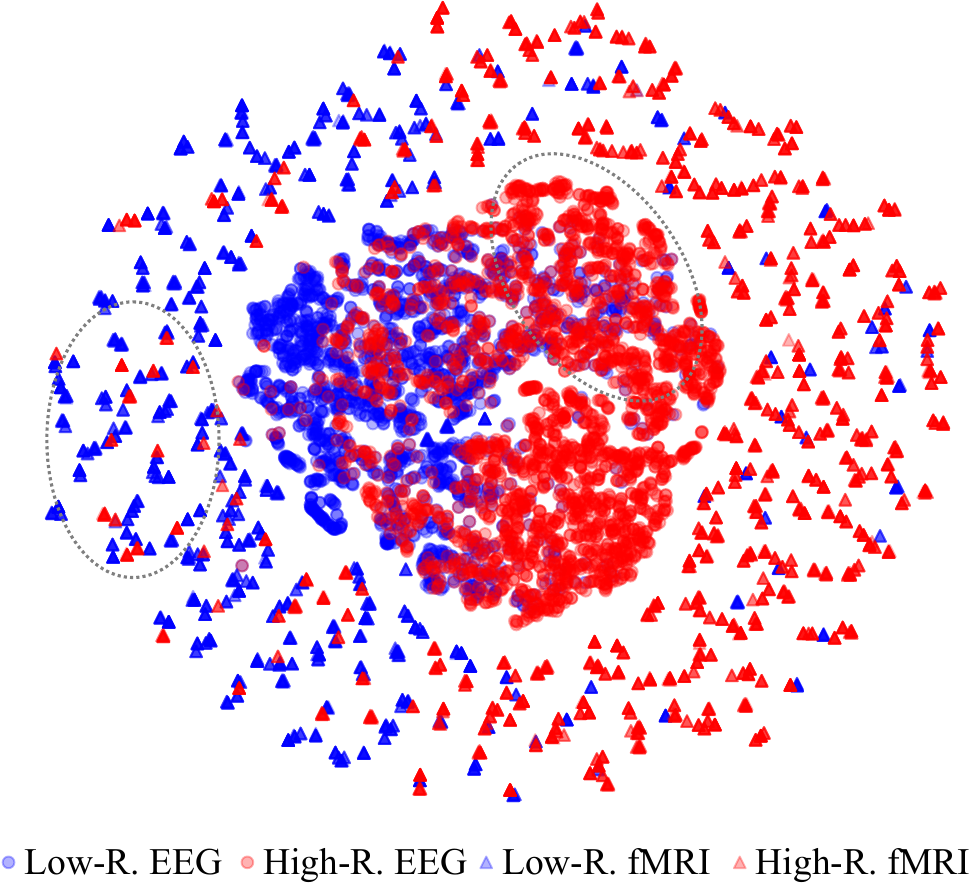}
    \end{center}
    \caption{2D visualization of EEG and fMRI features for low resilience and high resilience groups.}
    \label{figure:complementary}
\end{figure}

We conduct an ablation study to investigate the effectiveness of different designs in the proposed method. First, we ablate the Neuro Kolmogorov-Arnold Networks (KAN) operation and use multiple-layer perception with linear layers and ReLU activation functions for non-linear modeling. We then ablate the Triat-informed Multimodal Representation Learning (MRL) and replace it with the trivial CLIP loss for multimodal alignment. Finally, we remove the Noise-informed Prediction (NP) and use the chunk-level representation for classification and regression with the averaged ensemble, i.e., making predictions using chunk-level representations separately and ensembling the results as the final prediction.

As shown in Table \ref{table:ablation_study}, the performance drops on both datasets regardless of which design is ablated, showing that all modules contribute to the resilience prediction. We can find that ablating NP leads to the largest performance degradation on both datasets, suggesting without NP, the model fails to aggregate informative chunks and derive subject-level representations. Additionally, we observe that subject-aware contrastive loss contributes more to the LEMON Dataset. This might be due to the larger number of subjects in the LEMON dataset, where the impact of false negatives has a more pronounced effect on subject misidentification.

\subsection{Interpretability Experiments}

\begin{figure*}[]
    \begin{center}
    \includegraphics[width=0.95\linewidth]{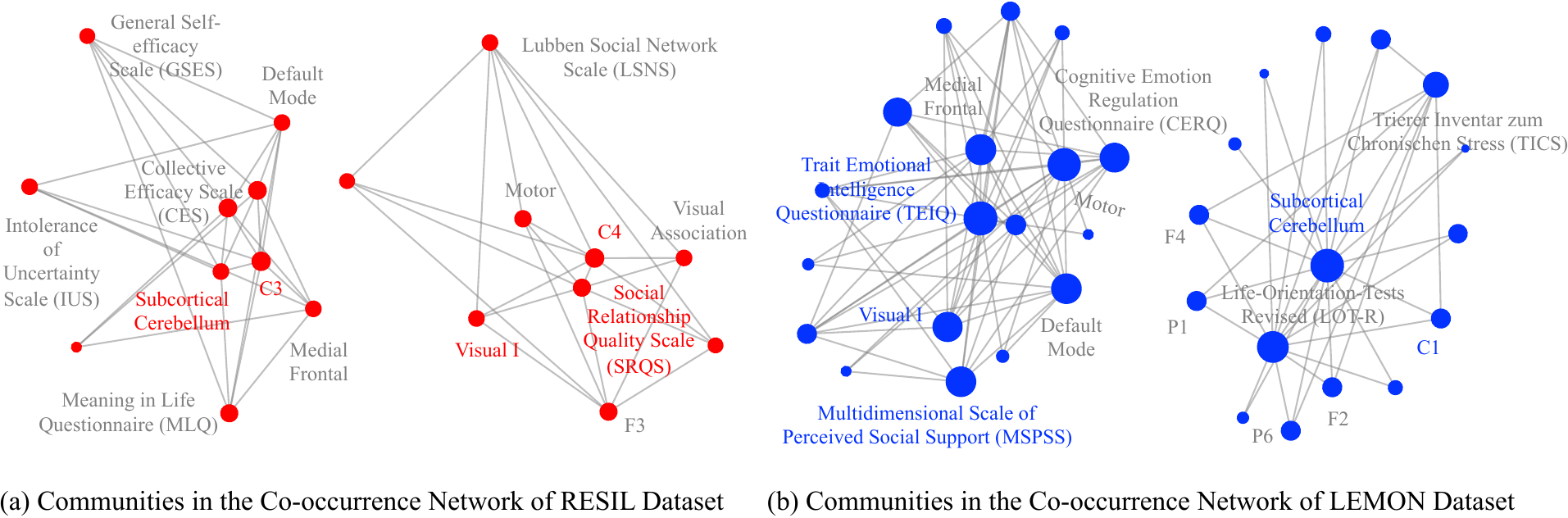}
    \end{center}
    \caption{Co-occurrence network of important regions in fMRI, electrodes in EEG, and scales in behavioral data based on the learned model of RESIL dataset.}
    \label{figure:cooccurance}
    \end{figure*}

\begin{figure}[]
    \begin{center}
    \includegraphics[width=\linewidth]{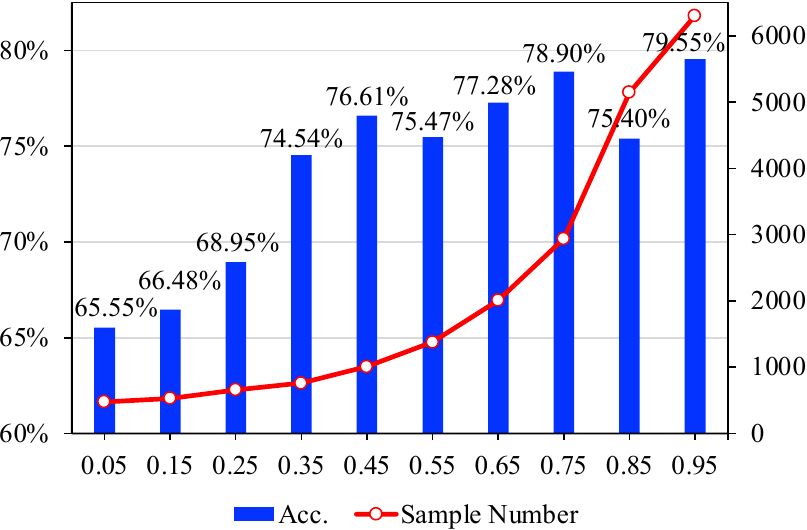}
    \end{center}
    \caption{Case study grouped by similarities on the RESIL dataset.}
    \label{figure:case_study}
\end{figure}
We then investigate the interpretability of the proposed method in psychological resilience. To locate informative patterns, we assign EEG and fMRI chunks into high and low-resilience groups. For EEG, we follow \cite{chung2024time} by adding perturbations to frequency bands at different electrodes and observing the resulting changes in multimodal similarity to derive importance scores. For fMRI analysis, due to numerous functional connections, we adopt guided backpropagation \cite{springenberg2014striving} to calculate the similarity between modalities and perform backpropagation, using the absolute gradient to derive importance scores. We then average these scores within groups to identify consistent important patterns in high and low-resilience subjects.

As shown in Fig. \ref{figure:eeg_fmri}(a), electrode C4 at the right lateral parietal cortex demonstrates high importance across theta and alpha bands in the low-resilience group. This aligns with previous studies \cite{quaedflieg2020stress,keunhoyoo2024power} showing decreased parietal activity at the right lateral parietal cortex during stress, with our results providing consistent evidence during sleep. Fig. \ref{figure:eeg_fmri}(c) reveals high importance of O2 across theta and alpha bands in the high-resilience group. While studies directly linking O2 to resilience mechanisms are limited, related research indicates that neural activity in the occipital cortex may regulate REM sleep \cite{wang2022rem}, suggesting unique sleep-stage characteristics. Additionally, Fig. \ref{figure:eeg_fmri}(b) and (d) show important functional connections between the inferior parietal lobule and the lateral temporal cortex, aligning with the default mode network's role in self-relevant decisions and affective cognitive processes \cite{miyagi2020psychological}.

\subsection{Exploration Experiments}

To gain further insights, we conduct experiments to explore the complementarity and consistency between modalities. For complementarity, we visualize EEG and fMRI representations on a 2D plane using t-SNE for the RESIL dataset samples. We use red and blue colors to indicate high and low resilience respectively, while circles represent EEG features and triangles represent fMRI features.

As shown in Fig. \ref{figure:complementary}, low resilience samples cluster on one side while high resilience samples cluster on the other, demonstrating successful alignment between EEG, fMRI, and behavioral data. Subjects with similar mental states are pulled closer, while those with different mental states are pushed apart. We observe that EEG distributions are concentrated but show overlap between low and high resilience clusters, while fMRI distributions are dispersed but show instance confusion. These characteristics demonstrate their complementary advantages.

To explore modality consistency, we analyze triplets of fMRI, EEG, and behavioral data. Using guided backpropagation \cite{springenberg2014striving}, we identify and record the most important regions, electrodes, and behavioral scales as co-occurring elements. We then record the frequency of different co-occurring elements and construct a co-occurrence network, where elements serve as nodes and co-occurrence frequencies as edges. We apply community detection after filtering the top 50\% co-occurring elements.

As shown in Fig. \ref{figure:cooccurance}, consistent co-occurrence patterns emerge across both datasets. Left parietal electrodes (C3, C1) co-occur with subcortical cerebellum regions and meaning-in-life scales, potentially indicating neural mechanisms related to internal factors. Additionally, Visual I regions co-occur with C4 (right parietal electrode) and social support scales, suggesting neural mechanisms linked to external factors.

\subsection{Case Study}

Finally, we conduct a case study to investigate the insights of NP, specifically examining why different chunks should be treated differently. We begin by forming EEG and fMRI chunk pairs from the same subject and iterating through these pairs. We bin the chunk pairs according to similarities between their multimodal representations and record the sample number across different similarity levels. We then evaluate classification performance using individual chunk pairs and analyze the performance distribution across different similarity levels.

As shown in Fig. \ref{figure:case_study}, we observe that the similarity follows a long-tail distribution, where most EEG and fMRI samples are similar, but some exhibit low similarity. This indicates that while most chunks are successfully aligned, some remain unaligned. Furthermore, we find that chunks with low alignment show poor classification accuracy (around 65\% when similarity is 0.05), suggesting these might be noisy chunks or chunks lacking resilience-related information. Based on these findings, we design our model to highlight aligned chunks while neglecting unaligned ones, achieving over 85\% accuracy with discriminative subject-level representations.

\section{Conclusions and Future Work}

This paper proposes a novel data-efficient model to predict resilience based on EEG and fMRI data. Several new techniques including a multimodal alignment algorithm to discover the nonlinear hidden patterns, a chucking algorithm for data argument, and a noise-informed inference algorithm to assess the quality of the neurological data, work together on the KAN-based architecture, and have achieved the impressive performance compared with SOTA algorithms on both RESIL and LEMON datasets. Interpretability experiments reveal learned patterns while exploration experiments show the complementarity and consistency between modalities. Recent work has also demonstrated the potential of KANs in providing interpretable symbolic formulas. In future work, we plan to leverage the interpretability of KANs to derive deeper neuropsychological insights.

\bibliographystyle{named}
\bibliography{ijcai25}

\begin{thebibliography}{}

\bibitem[\protect\citeauthoryear{Aghaomidi and Wang}{2024}]{aghaomidi2024ecg}
Poorya Aghaomidi and Ge~Wang.
\newblock Ecg-sleepnet: Deep learning-based comprehensive sleep stage
  classification using ecg signals.
\newblock {\em arXiv}, 2024.

\bibitem[\protect\citeauthoryear{Alter \bgroup \em et al.\egroup
  }{2024}]{alter2024robustness}
Tal Alter, Raz Lapid, and Moshe Sipper.
\newblock On the robustness of kolmogorov-arnold networks: An adversarial
  perspective.
\newblock {\em arXiv}, 2024.

\bibitem[\protect\citeauthoryear{Andrew \bgroup \em et al.\egroup
  }{2013}]{andrew2013deep}
Galen Andrew, Raman Arora, Jeff Bilmes, and Karen Livescu.
\newblock Deep canonical correlation analysis.
\newblock In {\em International Conference on Machine Learning}, pages
  1247--1255, 2013.

\bibitem[\protect\citeauthoryear{Babayan \bgroup \em et al.\egroup
  }{2019}]{babayan2019mind}
Anahit Babayan, Miray Erbey, Deniz Kumral, Janis~D Reinelt, Andrea~MF Reiter,
  Josefin R{\"o}bbig, H~Lina Schaare, Marie Uhlig, Alfred Anwander,
  Pierre-Louis Bazin, et~al.
\newblock A mind-brain-body dataset of {MRI}, {EEG}, cognition, emotion, and
  peripheral physiology in young and old adults.
\newblock {\em Scientific Data}, 6(1):1--21, 2019.

\bibitem[\protect\citeauthoryear{Benton \bgroup \em et al.\egroup
  }{2019}]{benton2019deep}
Adrian Benton, Huda Khayrallah, Biman Gujral, Dee~Ann Reisinger, Sheng Zhang,
  and Raman Arora.
\newblock Deep generalized canonical correlation analysis.
\newblock In {\em Association for Computational Linguistics}, 2019.

\bibitem[\protect\citeauthoryear{Bonanno}{2021}]{bonanno2021resilience}
George~A Bonanno.
\newblock The resilience paradox.
\newblock {\em European journal of Psychotraumatology}, 12(1):1942642, 2021.

\bibitem[\protect\citeauthoryear{Chung \bgroup \em et al.\egroup
  }{2024}]{chung2024time}
Hyunseung Chung, Sumin Jo, Yeonsu Kwon, and Edward Choi.
\newblock Time is not enough: Time-frequency based explanation for time-series
  black-box models.
\newblock In {\em International Conference on Information and Knowledge
  Management}, pages 394--403, 2024.

\bibitem[\protect\citeauthoryear{Connor and
  Davidson}{2003}]{connor2003development}
Kathryn~M Connor and Jonathan~RT Davidson.
\newblock Development of a new resilience scale: The connor-davidson resilience
  scale ({CD-RISC}).
\newblock {\em Depression and Anxiety}, 18(2):76--82, 2003.

\bibitem[\protect\citeauthoryear{Dong \bgroup \em et al.\egroup
  }{2018}]{dong2018stress}
Suh-Yeon Dong, Miran Lee, Heesu Park, and Inchan Youn.
\newblock Stress resilience measurement with heart-rate variability during
  mental and physical stress.
\newblock In {\em International Conference of the IEEE Engineering in Medicine
  and Biology Society}, pages 5290--5293, 2018.

\bibitem[\protect\citeauthoryear{Eldele \bgroup \em et al.\egroup
  }{2024}]{eldele2024tslanet}
Emadeldeen Eldele, Mohamed Ragab, Zhenghua Chen, Min Wu, and Xiaoli Li.
\newblock {TSLANet}: Rethinking transformers for time series representation
  learning.
\newblock In {\em International Conference on Learning Representations}, 2024.

\bibitem[\protect\citeauthoryear{Gao \bgroup \em et al.\egroup
  }{2020}]{gao2020connectome}
Mengxia Gao, Clive~HY Wong, Huiyuan Huang, Robin Shao, Ruiwang Huang,
  Chetwyn~CH Chan, and Tatia~MC Lee.
\newblock Connectome-based models can predict processing speed in older adults.
\newblock {\em NeuroImage}, 223:117290, 2020.

\bibitem[\protect\citeauthoryear{Gao \bgroup \em et al.\egroup
  }{2022}]{gao2022pyramidclip}
Yuting Gao, Jinfeng Liu, Zihan Xu, Jun Zhang, Ke~Li, Rongrong Ji, and Chunhua
  Shen.
\newblock {PyramidCLIP}: Hierarchical feature alignment for vision-language
  model pretraining.
\newblock {\em Advances in Neural Information Processing Systems},
  35:35959--35970, 2022.

\bibitem[\protect\citeauthoryear{Goel \bgroup \em et al.\egroup
  }{2022}]{goel2022cyclip}
Shashank Goel, Hritik Bansal, Sumit Bhatia, Ryan Rossi, Vishwa Vinay, and
  Aditya Grover.
\newblock {CyCLIP}: Cyclic contrastive language-image pretraining.
\newblock {\em Advances in Neural Information Processing Systems},
  35:6704--6719, 2022.

\bibitem[\protect\citeauthoryear{Hu \bgroup \em et al.\egroup
  }{2024}]{hu2024brainnpt}
Jinlong Hu, Yangmin Huang, Nan Wang, and Shoubin Dong.
\newblock {BrainNPT}: Pre-training transformer networks for brain network
  classification.
\newblock {\em Transactions on Neural Systems and Rehabilitation Engineering},
  2024.

\bibitem[\protect\citeauthoryear{Ingolfsson \bgroup \em et al.\egroup
  }{2020}]{ingolfsson2020eeg}
Thorir~Mar Ingolfsson, Michael Hersche, Xiaying Wang, Nobuaki Kobayashi, Lukas
  Cavigelli, and Luca Benini.
\newblock {EEG-TCNet}: An accurate temporal convolutional network for embedded
  motor-imagery brain--machine interfaces.
\newblock In {\em International Conference on Systems, Man, and Cybernetics},
  pages 2958--2965, 2020.

\bibitem[\protect\citeauthoryear{Jahin \bgroup \em et al.\egroup
  }{2024}]{jahin2024kacq}
Md~Abrar Jahin, Md~Akmol Masud, MF~Mridha, Zeyar Aung, and Nilanjan Dey.
\newblock {KACQ-DCNN}: Uncertainty-aware interpretable kolmogorov-arnold
  classical-quantum dual-channel neural network for heart disease detection.
\newblock {\em arXiv}, 2024.

\bibitem[\protect\citeauthoryear{Jiang \bgroup \em et al.\egroup
  }{2019}]{jiang2019removal}
Xiao Jiang, Gui-Bin Bian, and Zean Tian.
\newblock Removal of artifacts from {EEG} signals: a review.
\newblock {\em Sensors}, 19(5):987, 2019.

\bibitem[\protect\citeauthoryear{Jiang \bgroup \em et al.\egroup
  }{2024a}]{jiang2024large}
Weibang Jiang, Liming Zhao, and Bao-liang Lu.
\newblock Large brain model for learning generic representations with
  tremendous eeg data in bci.
\newblock In {\em International Conference on Learning Representations}, 2024.

\bibitem[\protect\citeauthoryear{Jiang \bgroup \em et al.\egroup
  }{2024b}]{jiang2024csp}
Xue Jiang, Lubin Meng, Xinru Chen, Yifan Xu, and Dongrui Wu.
\newblock {CSP-Net}: Common spatial pattern empowered neural networks for
  {EEG}-based motor imagery classification.
\newblock {\em Knowledge-Based Systems}, 305:112668, 2024.

\bibitem[\protect\citeauthoryear{Kalisch \bgroup \em et al.\egroup
  }{2017}]{kalisch2017resilience}
Raffael Kalisch, Dewleen~G Baker, Ulrike Basten, Marco~P Boks, George~A
  Bonanno, Eddie Brummelman, Andrea Chmitorz, Guill{\'e}n Fern{\`a}ndez,
  Christian~J Fiebach, Isaac Galatzer-Levy, et~al.
\newblock The resilience framework as a strategy to combat stress-related
  disorders.
\newblock {\em Nature Human Behaviour}, 1:784--790, 2017.

\bibitem[\protect\citeauthoryear{Kan \bgroup \em et al.\egroup
  }{2022}]{kan2022brain}
Xuan Kan, Wei Dai, Hejie Cui, Zilong Zhang, Ying Guo, and Carl Yang.
\newblock Brain network transformer.
\newblock {\em Advances in Neural Information Processing Systems},
  35:25586--25599, 2022.

\bibitem[\protect\citeauthoryear{KeunhoYoo \bgroup \em et al.\egroup
  }{2024}]{keunhoyoo2024power}
Kenny KeunhoYoo, Bowen Xiu, George Nader, Ariel Graff, Philip Gerretsen, Reza
  Zomorrodi, and Vincenzo De~Luca.
\newblock Power spectral analysis of resting-state eeg to monitor psychological
  resilience to stress.
\newblock {\em Psychiatry Research Communications}, 4(3):100175, 2024.

\bibitem[\protect\citeauthoryear{Keynan \bgroup \em et al.\egroup
  }{2019}]{keynan2019electrical}
Jackob~N Keynan, Avihay Cohen, Gilan Jackont, Nili Green, Noam Goldway,
  Alexander Davidov, Yehudit Meir-Hasson, Gal Raz, Nathan Intrator, Eyal
  Fruchter, et~al.
\newblock Electrical fingerprint of the amygdala guides neurofeedback training
  for stress resilience.
\newblock {\em Nature Human Behaviour}, 3(1):63--73, 2019.

\bibitem[\protect\citeauthoryear{Kipf and Welling}{2017}]{kipf2017semi}
Thomas~N. Kipf and Max Welling.
\newblock Semi-supervised classification with graph convolutional networks.
\newblock In {\em International Conference on Learning Representations}, 2017.

\bibitem[\protect\citeauthoryear{Klambauer \bgroup \em et al.\egroup
  }{2017}]{klambauer2017self}
G{\"u}nter Klambauer, Thomas Unterthiner, Andreas Mayr, and Sepp Hochreiter.
\newblock Self-normalizing neural networks.
\newblock {\em Advances in Neural Information Processing Systems}, 30, 2017.

\bibitem[\protect\citeauthoryear{Kong \bgroup \em et al.\egroup
  }{2015}]{kong2015neural}
Feng Kong, Xu~Wang, Siyuan Hu, and Jia Liu.
\newblock Neural correlates of psychological resilience and their relation to
  life satisfaction in a sample of healthy young adults.
\newblock {\em Neuroimage}, 123:165--172, 2015.

\bibitem[\protect\citeauthoryear{Krause \bgroup \em et al.\egroup
  }{2023}]{krause2023predicting}
Florian Krause, Judith van Leeuwen, Sophie Bogemann, Rayyan Tutunji, Karin
  Roelofs, Alex van Kraaij, Ruud van Stiphout, and Erno Hermans.
\newblock Predicting resilience from psychological and physiological daily-life
  measures.
\newblock {\em OSF Preprints}, 2023.

\bibitem[\protect\citeauthoryear{Lai \bgroup \em et al.\egroup
  }{2018}]{lai2018artifacts}
Chi~Qin Lai, Haidi Ibrahim, Mohd~Zaid Abdullah, Jafri~Malin Abdullah,
  Shahrel~Azmin Suandi, and Azlinda Azman.
\newblock Artifacts and noise removal for electroencephalogram ({EEG}): A
  literature review.
\newblock In {\em Symposium on Computer Applications \& Industrial
  Electronics}, pages 326--332, 2018.

\bibitem[\protect\citeauthoryear{Lau \bgroup \em et al.\egroup
  }{2021}]{lau2021integrative}
Way~KW Lau, Alan~PL Tai, Jackie~NM Chan, Benson~WM Lau, and Xiujuan Geng.
\newblock Integrative psycho-biophysiological markers in predicting
  psychological resilience.
\newblock {\em Psychoneuroendocrinology}, 129:105267, 2021.

\bibitem[\protect\citeauthoryear{Lawhern \bgroup \em et al.\egroup
  }{2018}]{lawhern2018eegnet}
Vernon~J Lawhern, Amelia~J Solon, Nicholas~R Waytowich, Stephen~M Gordon,
  Chou~P Hung, and Brent~J Lance.
\newblock {EEGNet}: a compact convolutional neural network for {EEG}-based
  brain--computer interfaces.
\newblock {\em Journal of Neural Engineering}, 15(5):056013, 2018.

\bibitem[\protect\citeauthoryear{Lazarus}{2020}]{lazarus2020psychological}
Richard~S Lazarus.
\newblock Psychological stress in the workplace.
\newblock In {\em Occupational Stress}, pages 3--14, 2020.

\bibitem[\protect\citeauthoryear{Lee \bgroup \em et al.\egroup
  }{2022}]{lee2022uniclip}
Janghyeon Lee, Jongsuk Kim, Hyounguk Shon, Bumsoo Kim, Seung~Hwan Kim, Honglak
  Lee, and Junmo Kim.
\newblock {UniCLIP}: Unified framework for contrastive language-image
  pre-training.
\newblock {\em Advances in Neural Information Processing Systems},
  35:1008--1019, 2022.

\bibitem[\protect\citeauthoryear{Li \bgroup \em et al.\egroup
  }{2021}]{li2021braingnn}
Xiaoxiao Li, Yuan Zhou, Nicha Dvornek, Muhan Zhang, Siyuan Gao, Juntang Zhuang,
  Dustin Scheinost, Lawrence~H Staib, Pamela Ventola, and James~S Duncan.
\newblock {BrainGNN}: Interpretable brain graph neural network for fmri
  analysis.
\newblock {\em Medical Image Analysis}, 74:102233, 2021.

\bibitem[\protect\citeauthoryear{Liu \bgroup \em et al.\egroup
  }{2025}]{liu2024kan}
Ziming Liu, Yixuan Wang, Sachin Vaidya, Fabian Ruehle, James Halverson, Marin
  Solja{\v{c}}i{\'c}, Thomas~Y Hou, and Max Tegmark.
\newblock {KAN}: Kolmogorov-arnold networks.
\newblock In {\em International Conference on Learning Representations}, 2025.

\bibitem[\protect\citeauthoryear{Liu}{2016}]{liu2016noise}
Thomas~T Liu.
\newblock Noise contributions to the {fMRI} signal: An overview.
\newblock {\em NeuroImage}, 143:141--151, 2016.

\bibitem[\protect\citeauthoryear{Manzoor \bgroup \em et al.\egroup
  }{2023}]{manzoor2023multimodality}
Muhammad~Arslan Manzoor, Sarah Albarri, Ziting Xian, Zaiqiao Meng, Preslav
  Nakov, and Shangsong Liang.
\newblock Multimodality representation learning: A survey on evolution,
  pretraining and its applications.
\newblock {\em Transactions on Multimedia Computing, Communications and
  Applications}, 20(3):1--34, 2023.

\bibitem[\protect\citeauthoryear{Miao \bgroup \em et al.\egroup
  }{2023}]{miao2023lmda}
Zhengqing Miao, Meirong Zhao, Xin Zhang, and Dong Ming.
\newblock Lmda-net: A lightweight multi-dimensional attention network for
  general eeg-based brain-computer interfaces and interpretability.
\newblock {\em NeuroImage}, 276:120209, 2023.

\bibitem[\protect\citeauthoryear{Miyagi \bgroup \em et al.\egroup
  }{2020}]{miyagi2020psychological}
Takashi Miyagi, Naoya Oishi, Kei Kobayashi, Tsukasa Ueno, Sayaka Yoshimura,
  Toshiya Murai, and Hironobu Fujiwara.
\newblock Psychological resilience is correlated with dynamic changes in
  functional connectivity within the default mode network during a cognitive
  task.
\newblock {\em Scientific Reports}, 10(1):17760, 2020.

\bibitem[\protect\citeauthoryear{Mu \bgroup \em et al.\egroup
  }{2022}]{mu2022slip}
Norman Mu, Alexander Kirillov, David Wagner, and Saining Xie.
\newblock {SLIP}: Self-supervision meets language-image pre-training.
\newblock In {\em European Conference on Computer Vision}, pages 529--544,
  2022.

\bibitem[\protect\citeauthoryear{Oh \bgroup \em et al.\egroup
  }{2024}]{oh2024geodesic}
Changdae Oh, Junhyuk So, Hoyoon Byun, YongTaek Lim, Minchul Shin, Jong-June
  Jeon, and Kyungwoo Song.
\newblock Geodesic multi-modal mixup for robust fine-tuning.
\newblock {\em Advances in Neural Information Processing Systems}, 36, 2024.

\bibitem[\protect\citeauthoryear{Paban \bgroup \em et al.\egroup
  }{2019}]{paban2019psychological}
Veronique Paban, Julien Modolo, Ahmad Mheich, and Mahmoud Hassan.
\newblock Psychological resilience correlates with {EEG} source-space brain
  network flexibility.
\newblock {\em Network Neuroscience}, 3(2):539--550, 2019.

\bibitem[\protect\citeauthoryear{Panicker and
  Gayathri}{2019}]{panicker2019survey}
Suja~Sreeith Panicker and Prakasam Gayathri.
\newblock A survey of machine learning techniques in physiology based mental
  stress detection systems.
\newblock {\em Biocybernetics and Biomedical Engineering}, 39(2):444--469,
  2019.

\bibitem[\protect\citeauthoryear{Quaedflieg \bgroup \em et al.\egroup
  }{2020}]{quaedflieg2020stress}
CWEM Quaedflieg, TR~Schneider, J~Daume, AK~Engel, and L~Schwabe.
\newblock Stress impairs intentional memory control through altered theta
  oscillations in lateral parietal cortex.
\newblock {\em Journal of Neuroscience}, 40(40):7739--7748, 2020.

\bibitem[\protect\citeauthoryear{Radford \bgroup \em et al.\egroup
  }{2021}]{radford2021learning}
Alec Radford, Jong~Wook Kim, Chris Hallacy, Aditya Ramesh, Gabriel Goh,
  Sandhini Agarwal, Girish Sastry, Amanda Askell, Pamela Mishkin, Jack Clark,
  et~al.
\newblock Learning transferable visual models from natural language
  supervision.
\newblock In {\em International Conference on Machine Learning}, pages
  8748--8763, 2021.

\bibitem[\protect\citeauthoryear{Samadi \bgroup \em et al.\egroup
  }{2024}]{samadi2024smooth}
Moein~E Samadi, Younes M{\"u}ller, and Andreas Schuppert.
\newblock Smooth kolmogorov arnold networks enabling structural knowledge
  representation.
\newblock {\em arXiv}, 2024.

\bibitem[\protect\citeauthoryear{Shen \bgroup \em et al.\egroup
  }{2013}]{shen2013groupwise}
Xilin Shen, Fuyuze Tokoglu, Xenios Papademetris, and R~Todd Constable.
\newblock Groupwise whole-brain parcellation from resting-state {fMRI} data for
  network node identification.
\newblock {\em Neuroimage}, 82:403--415, 2013.

\bibitem[\protect\citeauthoryear{Somvanshi \bgroup \em et al.\egroup
  }{2024}]{somvanshi2024survey}
Shriyank Somvanshi, Syed~Aaqib Javed, Md~Monzurul Islam, Diwas Pandit, and
  Subasish Das.
\newblock A survey on kolmogorov-arnold network.
\newblock {\em arXiv}, 2024.

\bibitem[\protect\citeauthoryear{Song \bgroup \em et al.\egroup
  }{2022}]{song2022eeg}
Yonghao Song, Qingqing Zheng, Bingchuan Liu, and Xiaorong Gao.
\newblock {EEG} conformer: Convolutional transformer for {EEG} decoding and
  visualization.
\newblock {\em Transactions on Neural Systems and Rehabilitation Engineering},
  31:710--719, 2022.

\bibitem[\protect\citeauthoryear{Springenberg \bgroup \em et al.\egroup
  }{2014}]{springenberg2014striving}
Jost~Tobias Springenberg, Alexey Dosovitskiy, Thomas Brox, and Martin
  Riedmiller.
\newblock Striving for simplicity: The all convolutional net.
\newblock {\em arXiv}, 2014.

\bibitem[\protect\citeauthoryear{Steyer \bgroup \em et al.\egroup
  }{1999}]{steyer1999latent}
Rolf Steyer, Manfred Schmitt, and Michael Eid.
\newblock Latent state--trait theory and research in personality and individual
  differences.
\newblock {\em European Journal of Personality}, 13(5):389--408, 1999.

\bibitem[\protect\citeauthoryear{Sturzbecher and de
  Araujo}{2012}]{sturzbecher2012simultaneous}
Marcio~Junior Sturzbecher and Draulio~Barros de~Araujo.
\newblock Simultaneous {EEG-fMRI}: integrating spatial and temporal resolution.
\newblock {\em The Relevance of the Time Domain to Neural Network Models},
  pages 199--217, 2012.

\bibitem[\protect\citeauthoryear{Tang \bgroup \em et al.\egroup
  }{2024}]{tang20243d}
Tianze Tang, Yanbing Chen, and Hai Shu.
\newblock 3d u-kan implementation for multi-modal mri brain tumor segmentation.
\newblock {\em arXiv}, 2024.

\bibitem[\protect\citeauthoryear{Wang \bgroup \em et al.\egroup
  }{2022}]{wang2022rem}
Ziyue Wang, Xiang Fei, Xiaotong Liu, Yanjie Wang, Yue Hu, Wanling Peng,
  Ying-wei Wang, Siyu Zhang, and Min Xu.
\newblock Rem sleep is associated with distinct global cortical dynamics and
  controlled by occipital cortex.
\newblock {\em Nature Communications}, 13(1):6896, 2022.

\bibitem[\protect\citeauthoryear{Wang \bgroup \em et al.\egroup
  }{2024}]{wang2024generalizable}
Jiquan Wang, Sha Zhao, Haiteng Jiang, Shijian Li, Tao Li, and Gang Pan.
\newblock Generalizable sleep staging via multi-level domain alignment.
\newblock In {\em AAAI Conference on Artificial Intelligence}, volume~38, pages
  265--273, 2024.

\bibitem[\protect\citeauthoryear{Watanabe}{2023}]{watanabe2023tree}
Shuhei Watanabe.
\newblock Tree-structured parzen estimator: Understanding its algorithm
  components and their roles for better empirical performance.
\newblock {\em arXiv}, 2023.

\bibitem[\protect\citeauthoryear{Xiang \bgroup \em et al.\egroup
  }{2024}]{xiang2024using}
Fayang Xiang, Li~Zhang, Yidan Ye, Chuyue Xiong, Yanjie Zhang, Yan Hu, Jiang Du,
  Yi~Zhou, Qiyue Deng, and Xinke Li.
\newblock Using pupil diameter for psychological resilience assessment in
  medical students based on {SVM} and {SHAP} model.
\newblock {\em Journal of Biomedical and Health Informatics}, 2024.

\bibitem[\protect\citeauthoryear{Xue \bgroup \em et al.\egroup
  }{2023}]{xue2023ulip}
Le~Xue, Mingfei Gao, Chen Xing, Roberto Mart{\'\i}n-Mart{\'\i}n, Jiajun Wu,
  Caiming Xiong, Ran Xu, Juan~Carlos Niebles, and Silvio Savarese.
\newblock {ULIP}: Learning a unified representation of language, images, and
  point clouds for 3d understanding.
\newblock In {\em Computer Vision and Pattern Recognition}, pages 1179--1189,
  2023.

\bibitem[\protect\citeauthoryear{Zhang and
  Zhou}{2024}]{zhang2024generalization}
Xianyang Zhang and Huijuan Zhou.
\newblock Generalization bounds and model complexity for kolmogorov-arnold
  networks.
\newblock {\em arXiv}, 2024.

\end{thebibliography}

\end{document}